
\input phyzzx
\def\CWRU{Dept. of Physics, Case Western Reserve University,
Cleveland, OH   44106-7079}
 \PHYSREV
 \doublespace
 \pubnum{92-6}
 \date{June 1992}
 \pubtype{T/E}
 \titlepage
\hoffset=.3125in
\voffset=.3125in
 \title{Disoriented Chiral Condensates:\break
 A White Paper for the Full Acceptance Detector
}
 \author{K.~L.~Kowalski and C.~C.~Taylor}
 \CWRU
\abstract
\singlespace
Theoretical speculations and experimental data suggesting the
possibility of observing disoriented chiral condensates at a Full Acceptance
Detector are reviewed.
\endpage

 \sequentialequations

\chapter{Introduction}

These notes are intended as an introduction to the theory and phenomenology
of disoriented chiral condensates.  Briefly speaking, we are concerned with
the possibility that coherent states of pions might be produced in
hadron-hadron collisions in existing or planned particle accelerators.  Such
phenomena would be signaled by events with anomalously large (or small)
numbers of $\pi^0$'s in comparison with the number of charged pions.  One
purpose of these notes is to attempt to estimate the cross section for
production of such events, the expected distribution of the neutral pion
frequency, and the expected momentum distribution of such events.  In the
process of developing these estimates we will review theoretical grounds
for believing that such events may occur.  We shall also attempt to
confront these theoretical ideas with available data, concentrating on
the phenomenology of Centauro-type events.  We hope that before reaching
the end
of these notes the reader agrees with us that this is an area which is
fascinating both theoretically and experimentally, and that the interest
is such as to warrant mounting additional efforts to observe the
associated phenomena.

Before beginning, we note that most of the basic ideas contained here
will be found in the various papers Bjorken has written in the past year
in support of a Full-Acceptance Detector\REFS\refbjone{J.D. Bjorken,
``A Full Acceptance Detector for SSC Physics at Low and Intermediate
Mass Scales", (SSC EoI-19), SLAC-PUB-5545 (May 1991).}
\REFSCON\refbjtwo{J.D. Bjorken, SLAC-PUB-5673.}
\REFSCON\refbjthree{J.D. Bjorken and M. Weinstein, private communication}
\REFSCON\refbjfour{J.D. Bjorken, ``How Black is a Constituent Quark",
SLAC-PUB-5794 (April 1992).
}\refsend.
This present work is mainly
designed to try to develop the ideas a bit further, and attempt to
confront the ideas with the occurrence of Centauro events and
related phenomena.

First, what is a ``disoriented chiral condensate?"  This requires a brief
digression into the physics of chiral symmetry.

Consider QCD with
an isospin doublet of massless quarks
$$\psi = \pmatrix{u\cr d\cr}.$$
The fermion terms in the Lagrangian are invariant under both global
isospin transformations
$$\delta \psi = i \epsilon\cdot T \psi$$
and under global chiral transformations
$$ \delta \psi = i \gamma_5 \epsilon\cdot T \psi.$$
It is widely assumed that the chiral symmetry is spontaneously broken in
the QCD ground state, leading to the interpretation of the pion as a massless
Goldstone boson, and generating a mass for the quarks in the process.
(In the real world, there are, apparently, small quark masses explicitly
breaking chiral symmetry in the QCD Lagrangian, leading to the small but
non-vanishing pion mass.)  A signal of the
spontaneous breaking of chiral symmetry
is the non-vanishing of the  vacuum
condensate $\langle {\bar\psi} \psi \rangle$.

The physics which we are studying here corresponds to the possibility
that in a macroscopic (but localized) region of spacetime, the vacuum
condensate may be disoriented from its external or ambient value,
so that the condensate is associated with
the isovector, rather than the isoscalar, degrees of freedom.

Our first theoretical problem is one of a choice of language.  While we could,
in principle, work in the full theory of QCD, this appears to be much
too complicated.  Instead, since at least the late stages of the evolution
of the chiral condensates are expected to involve characteristic momenta which
are small compared to $\Lambda_{QCD}$, we find it more convenient to discuss
the phenomena in terms of a low energy effective theory:  the sigma model.
At various times in our discussion we will be faced with the question of
which variation of the sigma model is appropriate.  For example, if we
want to consider intermediate states in which a region of spacetime is
locally in a chirally symmetric phase, then we will adopt the linear
sigma model.  We will also from time to time want to include quarks in
our discussion, in which case we shall resort to the chiral quark model,
in either it's linear or non-linear incarnation, as appropriate.  We should
note at the outset that the use of the non-linear sigma model to calculate
low energy processes is well established\Ref\don{A recent brief summary is
J.~F. Donoghue, ``Introduction to Nonlinear Effective Field Theory,"
UMHEP-360, 1992.}.  Incorporating quarks in this
version of the model (the chiral quark model) also has some successes, but
is beset by questions relating to the potential double-counting of the
pion\Ref\georman{
H.~Georgi and A.~Manohar, {\sl Nucl. Phys.} {\bf B234} 189 (1984).}.
In discussions of the chiral phase transition, the utility of the linear
sigma model has again been recently urged\REFS\wilczone{F. Wilczek,
``Application of
the Renormalization Group to Second Order QCD Phase Transition",
IASSNS-HEP-91/65 (January 1992).}
\REFSCON\wilcztwo{F. Wilczek, ``Remarks on the
phase transition in QCD", IASSNS-HEP-92/23 (March 1992).}
\refsend ; the validity of the linear, chiral
quark model has also been argued\Ref\goksch{A. Goksch, {\sl Phys. Rev.
Lett.} {\bf 67} 1701 (1991).}, although this appears to be
rather controversial\Ref\bernard{C. Bernard, et al., {\sl Phys. Rev. Lett.}
{\bf 68} 2125 (1992).}.
Finally, on occasion we will include explicit chiral symmetry breaking
in the form of a non-vanishing pion mass.

The upshot of all of this is that we appear to have few quantitative
theoretical tools at our disposal, particularly in addressing questions
related to the production of disoriented chiral condensates.  Nevertheless,
we do have more or less reliable qualitative and semi-quantitative tools
for describing disoriented chiral condensates, particularly in the late
stages of their evolution.

At this point, it may be useful to introduce
a simple model for the production of disoriented chiral
condensates.  In this picture, we imagine that in some fraction of the
collisions between `constituent quarks' at sufficiently high energies,
a significant
fraction of the incident momentum is thermalized, resulting in small
regions of extremely hot hadronic matter, presumably hot enough to be
in a chirally symmetric phase.  We then expect the hadronic debris to
expand outwards (transverse flow is important!) at a speed approaching
that of light leaving the interior
to cool rapidly.  As the interior thus drops below the critical
temperature, it will undergo a phase transition to a broken phase.  What
is crucial for our purposes is that the order parameter characterizing
this phase need not have the same value that it does in the rest of the
universe since the interior is protected from the exterior by a hot
shell of rapidly expanding hadronic matter that is presumably in a
chirally symmetric phase.  Eventually,
the shell cools to the hadronization temperature, and the interior
re-establishes contact with the rest of the universe.  At this point,
the interior will realign with the exterior through the coherent
radiation of non-relativistic pions.  It is amusing to note the similarities
between this picture and the `Baked Alaska' model for nucleation of phase
transitions in superfluid He-3 \Ref\bakal{{\sl Phys. Today}, {\bf 45} \#2,
20 (June 1992).}.

We consider the preceding scenario to be suspect in many regards.  It is
far from clear that such a classical picture is appropriate,
particularly in the early phases.  Indeed, it is not clear that the process
of creating disoriented chiral condensates need even proceed via an
intermediate state dominated by a localized symmetric vacuum state.
Nevertheless we believe that it is a useful starting point for discussing
some of the anticipated phenomenology of disoriented chiral condensates.

	In this context, a working definition of an event involving a
disoriented chiral condensate is:

\item{(a)} At a late stage in the production process,
just before the relaxation
of the inside 	to the outside vacuum, there is (in a suitable frame) a
roughly spherical	volume of disoriented chiral condensate enveloped by a
uniformly packed shell 	of hadrons that are moving outward at the speed
of light.

\item{(b)}The localized chiral condensate,
(the ``inside vacuum") then relaxes
to the 	outside vacuum accompanied by the radiation of Goldstone bosons,
viz., primarily 	pions.

The relevant physics then involves three fundamental issues:

\item{(i)} Is there a threshold realizable at existing or
currently planned hadron colliders for producing a disoriented chiral
condensate and
what are 	testable models of possible production mechanisms in pp
collisions?

\item{(ii)} What are the orders of magnitude of the cross sections for the
occurrence of 	processes in which the disorientation and subsequent
relaxation of the chiral condensate plays a central role?

\item{(iii)} What are the expected experimental signals for such processes?

Theoretical insight on these questions is presently rudimentary because the
underlying processes are both dynamic (nonstatic) and nonperturbative.
Progress in the analysis of disoriented chiral condensate-generated physics
and properties will be
strongly driven by an F.A.D. experimental program.

We shall address issues (i) and (ii) below; for the moment, we turn to
the question of experimental signals of disoriented chiral condensates.

\chapter{A First Look at Experimental Signatures}

We can use the physical picture of the late stages of
the evolution of a disoriented chiral
condensate developed in the previous section to develop a semi-quantitative
understanding of some of the experimental signatures of such objects.  We
shall begin with an essentially classical picture, and then refine the
picture to incorporate obvious quantum-mechanical effects.

We  use the  sigma model, without quarks, to
model the chiral dynamics.  In particular, we assume that the dynamical system
can be described by the Lagrangian
$${\cal L} = {1\over2}\, ( \partial_\mu {\vec \pi}\, \partial^\mu {\vec \pi} +
\partial_\mu \,\sigma \partial^\mu \sigma) -
{\lambda\over 2}\, ( {\vec \pi}^2 + \sigma^2 -f_\pi^2)^2.
$$

The minimum of the potential is at $ \sigma^2 + {\vec \pi}^2 = f_\pi^2$,
so that the chiral  symmetry is spontaneously broken by having
$\langle \sigma \rangle = f_\pi$.  Since $\sigma$ is a scalar, this is
the translation into the language of the effective theory of the belief
that the scalar/isoscalar quark bilinear develops a vacuum expectation
value in the full theory of QCD.

The dynamics we are interested in  simplifies in the non-linear
limit of $\cal L$.  That is, either $\lambda$
is a Lagrange multiplier enforcing the constraint ${\vec \pi}^2 + \sigma^2
=f_\pi^2$, or, equivalently, $\lambda \to \infty$.  This non-linear limit
seems to be a reasonable approximation: If one solves for the $\sigma$ mass
using the tree-level Lagrangian, one finds
$ \lambda \approx m_\sigma^2 /(4 f_\pi^2)>>1$.  One-loop corrections reduce
this, but still leave $\lambda \sim 6$\Ref\lee{B. W. Lee,
{\sl Chiral Dynamics},
Gordon and Breach, 1972.}.  We have explicitly checked some
aspects of our conclusions through numerical solutions at finite $\lambda$;
the conclusions are largely insensitive to this approximation.

With these assumptions we can  model the relaxation process by studying
the classical evolution
of a domain-wall-like field configuration, suitably interpreting the eventual
outgoing fields as quantum mechanical coherent states.  The physical picture
outlined above suggests that we assume spherical symmetry and begin  our study
of the dynamics at the time of
hadronization of the spherical debris shell. Specifically, we assume that
in the interior, the chiral condensate is disoriented by an angle $\theta_0$
from the $\sigma$ direction, while in the exterior the chiral condensate
is composed only of the $\sigma$ field.
Thus, it seems reasonable to assume that the
fields satisfy
$$\eqalign{{\vec \pi}(r,0)&\to 0\;{\rm as}\; r\to \infty,\cr
          {\vec \pi(r,0)}&\to
 {\hat n}f_\pi\sin{(\theta_{0})} \; {\rm as}\; r\to 0,\cr
           \sigma(r,0)&\to f_\pi\;{\rm as}\; r\to \infty,\cr
           \sigma(r,0)&\to f_\pi\cos{(\theta_{0})} \; {\rm as}\; r\to 0.}
$$
In these equations $\hat n$ is an arbitrary unit vector in isospin space.

Now, in order to study the dynamics, we need to make specific assumptions
about the profile of both the fields and their velocities.  We assume
that the fields at time $t=0$ can be modeled by a stationary
kink-like configuration.  It is easiest to express this by writing
$$\eqalign{ {\vec\pi} &= f_\pi \sin{\theta(r,t)}, \cr
\sigma&= f_\pi \cos{\theta(r,t)}},$$
with, for example,
$$\theta(r,t=0) = \theta_0 (1-\tanh{({r-R\over T})}),$$
with $\dot\theta=0$ and obvious candidate for a model of the interface.

There are several point to be made.

First, in this idealization of the
dynamics, the evolution of the system is determined by an effective
Lagrangian for $\theta$, which is just
$${\cal L}_{\rm eff} = {f_\pi^2\over 2} (\partial_\mu \theta)^2.$$
That is, the later stages of evolution are essentially that of a free
field determined by the configuration of the disoriented condensate
at the time of decoupling.  This observation emphasizes that (in this
scenario) the thickness of the interface must be determined by the
thickness of the `hot' shell of hadronic matter at the time of
hadronization.  This will permit us to make some semi-quantitative
estimates of particle production.

Second, in this idealization (with a massless pion),
the only contribution to the energy of the disoriented condensate comes
from the interface between the interior and the exterior.  We shall later
generalize this to include the volume-dependent contribution arising
from a non-vanishing pion mass.

How do we interpret this classical pion field?  The obvious way of
making a connection with quantum field theory is to interpret
the outgoing wave as a coherent state.
Specifically,  one might consider
a coherent state
$$ |\eta\rangle =
e^{\int d^3k { a}^\dagger(k) \eta(k)}|0\rangle$$
with the property that
$$a(k)|\eta  \rangle =\eta(k)|\eta \rangle,$$
where $$\langle \eta  | \eta  \rangle =
e^{(\int d^3k |\eta(k)|^2)}.$$
The argument of the exponential is interpreted as the mean number of
particles.  (Note that we have temporarily suppressed isospin indices.)

There is, however, a subtlety associated with the interpretation in terms
of coherent states which is extremely important both in the discussion of
experimental signatures and in developing a deeper quantum-mechanical
understanding of the disoriented chiral condensates.  The issue involved
is easily uncovered by the following series of considerations.  First,
we can consider a disoriented condensate in a specific cartesian isospin
direction  $\hat n$.  If, in our classical discussion, both
$\vec\pi$ and $\dot{\vec\pi}$ are in the $\hat n$ direction, then the
field configuration has vanishing isospin density, and hence, vanishing
total isospin.  This is easily seen by noting that the isospin generators
are given by
$$Q^a \sim \int {\rm d^3 x}\;{\vec \pi} \times {\dot{\vec\pi}}.$$
However, the coherent states described above present us with a paradox.
Consider a chiral condensate disoriented in, say, the $\pi^1$ direction.
The coherent state will then be of the form
$$ |\eta\rangle = e^{\int d^3k a_1^\dagger(k) \eta_1(k)}|0\rangle.$$
But $a_1^\dagger = 1/\sqrt{2} \;(a_+^\dagger +a_-^\dagger)$, so that the
coherent state will, for instance, have non-vanishing matrix elements
with states of arbitrarily large charge!

The resolution to this is straightforward.  All directions $\hat n$ are
equally accessible, and so quantum mechanically one should average over
them.  This is equivalent to making a projection on the isospin-zero
sector of the coherent states described above.  This in turn leads us to
the most striking experimental signal of the production of disoriented
chiral condensates:  a distribution of the fraction $f$ of
neutral pions $\sim 1/\sqrt{f}$.

We now give a simple derivation of the predicted isospin
distribution of events with large number of pions, under the assumptions
that the state is an isospin singlet,  and that all of the pions have the
same spatial wave function.  The formulae we shall derive are valid for
any number of pions in such a state.  In the limit that the number of
pions is large, we shall find that the distribution expected is $\sim 1/
\sqrt{f}$, where $f$ is the fraction of neutral pions.

Since we are concerned with $2 N$ identical pions, we can carry out the
derivation by introducing the operators $a^\dagger_i, a_i$
which create/annihilate pions in the relevant state.  (The  index `i' is
a cartesian isospin index).  We take these operators to be normalized
so that $\lbrack a_i, a^\dagger_j \rbrack = \delta_{i  j}$.  The isospin
generators are then realized by $I_i = - i \epsilon_{ijk} a^\dagger_j a_k$.
While classically we shall find it useful to work in the cartesian isospin
basis, quantum mechanically it is more convenient to work in a basis in
which $I_3$ is diagonal, and introduce appropriate raising and lowering
operators.  It is thus more convenient to work in terms of the creation
operators $a_3, a^\dagger_\pm =
1/\sqrt{2}(a^\dagger_1 \pm i a^\dagger_2)$ and their hermitian
conjugates.  Similarly, it is convenient to work with the z-component
of the isospin generators
$$I_3 =  a_+^\dagger a_+ - a_-^\dagger a_-$$
and the corresponding isospin raising/lowering operators
$$I_\pm = \pm {\sqrt{2}} ( a^\dagger_\pm a_3 + a_3^\dagger a_\mp).$$

We now turn to the construction of a multi-pion state ``coherent"
state $|\psi\rangle$  which
is an isosinglet and an eigenstate of the total pion number operator.
Requiring $I_3 |\psi\rangle =0$ implies that $n_+=n_-$, and hence that
the total number of pions is even,
as is the number of neutral pions.
Thus we can expand
$$|\psi\rangle = \sum_{n=0}^N C^{(N)}_n (a^\dagger_3)^{2 n} (a_+^\dagger
 a_-^\dagger)^{N-n}|0\rangle,$$
where $2 N$ is the total number of pions, $2 n$ is the number of $\pi^0$'s,
$|0\rangle$ is annihilated by the $a_i$, and the $C^{(N)}_n$ are expansion
coefficients to be determined by the requirement that $I_\pm |\psi\rangle=0$.

We can determine the expansion coefficients up to an overall normalization
and phase fairly straightforwardly.  First, observe that requiring
$I_+  |\psi\rangle=0$ gives us $N$ linear equations for the $N+1$ unknown
 $C^{(N)}_n$.  (The $I_-$ equation yields no additional information.) These
equations can be solved in terms of a single coefficient, say $C^{(N)}_0$.

We can also by inspection find an explicit form for a $2N$ pion isosinglet
state as follows. The operator
$$S^\dagger= 2 a_+^\dagger a_-^\dagger - (a_3^\dagger)^2$$
commutes with the isospin generators, $\lbrack I_i, S^\dagger \rbrack =0$.
Consequently, the state
$$|\psi\rangle = C^{(N)}_0 (2 a_+^\dagger a_-^\dagger - (a_3^\dagger)^2)^N
 |0\rangle$$
is a $2N$ pion isosinglet, and, by the considerations of the preceeding
paragraph, must be the most general such state in which all of the pions
have the same spatial wave function.

{}From this it is possible to determine the probability for seeing
$2n$ neutral pions out of $2 N$ total pions in such a state:
$$P(n,N) = {(N!)^2 2^{2N}\over (2N+1)!} {(2n)!\over{(n! 2^n)^2}}$$
Using Stirling's approximation, it is straightforward to demonstrate that
$P(n,N)\sim 1/\sqrt{n/N}$ in the limit that both $n$ and $N$ become large.
These results have also been previously found using other
arguments \refmark\refbjone,
\REFS\hornsilv{D. Horn and R. Silver, {\sl Ann. Phys. (N.Y.)}
{\bf 66} 509 (1971).}
\REFSCON\eriksson{K.-E. Eriksson, N. Mukunda and B.-S. Skagerstam,
{\sl Phys. Rev.}
{\bf D24} 2615 (1981).}
\REFSCON\ansrysk{A. A. Anselm and M. G. Ryskin, {\sl Phys. Lett.} {\bf B266}
482
(1991).}
\REFSCON\bk{J.-P. Blaizot and A. Krzywicki, ``Soft Pion Emission in Heavy-Ion
Collisions", LPTHE Orsay 92/11.}
\refsend.

\chapter{Tentative Phenomenology}

Having established the striking signature of the $1/\sqrt{f}$ distribution
of neutral pions, it is appropriate to try to develop the expected
phenomenology in a little more detail.  We are hampered a bit in that
we cannot do any calculations from first principles.  Nevertheless, we
can make a number of estimates which should be useful guides in searching
for disoriented chiral condensates.  Similar estimates have been presented
by Bjorken\refmark\refbjfour.

We begin, as before, from the physical picture developed in the
introductory section.  That is, we consider a spherical region of
radius $R$ within which the chiral condensate is disoriented by an
angle $\theta_0$ from the sigma direction.  The interior is separated
from the exterior by an interface of thickness $T$ which is determined
by the dynamics of the hot shell of hadronic matter resulting from the
initial collision.  We first estimate the energy density of the configuration
at the time of hadronization of the shell.  In the interior, we assume
that there is essentially no kinetic energy, but there is an energy
density arising from the non-vanishing pion mass:
$$ {\cal E}_V \approx {1\over 2} m_\pi^2 {\vec \pi}^2 \approx
 {1\over 2} m_\pi^2 f_\pi^2 \sin^2{(\theta_0)}.$$
Similarly, there will be an energy density associated with the interface
between the interior and the exterior.  The dominant contribution in
this region will be that due to the gradients of the fields:
$${\cal E}_S \approx {1\over2}\, ( \sum_a (\nabla \pi_a)^2 +
(\nabla \sigma)^2)\approx 2 {f_\pi^2\over T^2} \sin^2{(\theta_0/2)}.$$
The corresponding total energies are thus
$$E_V \approx {2\over 3} \pi R^3 m_\pi^2 f_\pi^2 \sin^2{(\theta_0)}$$
and
$$E_S \approx 8 \pi {R^2\over T} f_\pi^2 \sin^2{(\theta_0/2)}.$$

To proceed further, we need to estimate some of the relevant parameters.
The shell cannot be thinner, at the time of hadronization, than the
pion radius $r_\pi \approx 2/3 \;fm$.
We take this to be our best estimate for $T$. Thus, if we assume
a dense layer of pions of this thickness, we have a total of
$$N_{nor}\approx {4  R^2\over r_\pi^2}.$$  Note that it seems
reasonable to assume that these ``normal" pions have typical average
transverse momentum
$\langle p_t \rangle \approx
.5 \;GeV$.  We note that this model predicts that ``normal" multiplicty
should scale with the surface area of the sphere.  There is some evidence
for this from studies of pion interferometry.  These appear to be broadly
consistent with the picture we are assuming, though details of the comparison
are complicated by details of the data analysis.  It is not clear to us
whether the data favors a larger or smaller value for $T$.

Our estimate of $T\approx 2/3\; fm$ implies that the pions associated with
the surface energy will have characteristic momenta of the order
$p_S \sim 2/T\approx .6\; GeV$.  Thus, this component of the pion
spectrum will be relativistic.  Hence, we estimate the number of pions
associated with the interface energy as
$$N_S \approx{ E_S\over p_S}
 \approx 4 \pi {R^2} f_\pi^2 \sin^2{(\theta_0/2)}.$$
Similarly, assuming that $R>>T$, so that the physical picture outlined above
is meaningful, the pions associated with the interior volume will be
non-relativistic.  Hence the total number of pions associated with the
interior condensates is
$$N_V \approx {E_V\over m_\pi} \approx
{2\over 3}\pi f_\pi^2 m_\pi R^3 \sin^2{(\theta_0)}.$$

While we cannot estimate thresholds in the absence of a tractable
dynamical model of the early stages, we can define criteria for our
picture to be self consistent.  Specifically, the picture that we
have suggested implicitly assumes that the system be macroscopic, with
$R>>T$.  Taking $R\sim 5 T \sim 3\; fm$ seems a criterion for a
`threshold' above which our picture may be sensible.  Taking $f_\pi=93\;
MeV$ we thus estimate at this `threshold':
$N_{nor} \approx 83$ pions, $N_S\approx 24.5 \sin^2{(\theta_0/2)}$ pions, and
$N_V \approx 8.6 \sin^2{(\theta_0)}$ pions.  Such an object would have
an energy in its center of mass frame of approximately
$$M^* \approx  E_{nor} + E_S + E_V \approx 41.5\; GeV\; +
16.0 \sin^2{(\theta_0/2)} \; GeV\; + 1.3 \sin^2{(\theta_0)}\; GeV.$$
If we assume that
 $\langle \sin^2{(\theta_0/2)}\rangle =
\langle \sin^2{(\theta_0/2)}\rangle =1/2$, which is probably reasonable
as long as the energy of the condensate is small compared to the energy
of the normal shell, we have $M^*\approx 50\; GeV$ carried by $\sim 100$
pions.

As the size of the object increases, the importance of the chiral condensate
increases because of the volume contribution. However, the coefficient of
the volume term is rather small, so that it will not be the dominant
contribution until $ R\sim 28\; fm$, corresponding to pion multiplicities
of $\sim 4000$!

These estimates raise one issue which may be extremely important for
the phenomenology.  In our discussion we have divided the system at
the time of
hadronization into three parts:  a surface layer of `normal' pions, an
interior disoriented chiral condensate, and an interface between the
interior condensate and the exterior vacuum whose thickness is controlled
by the `normal' pions.  Both the interface and the interior condensate
contribute to the production of coherent pions, with the distinction
serving mainly to suggest the momentum distribution of the condensate.
One needs to worry, however, that the distinction between the `normal'
pions (which scale like $R^2$) and the interface contribution to the
coherent pions (which also scales like $R^2$) is overdrawn. Our estimates
suggest that they have comparable momentum scales, the ratio of the
number of coherent pions from the interface energy to the number of
normal pions is estimated to be
$${N_S\over N_{nor}} \approx \pi f_\pi^2 r_\pi^2 \sin^2{(\theta_0/2)}\approx
0.3 \sin^2{(\theta_0/2)}.$$
While for small values of $\theta$ the distinction is probably reasonable,
one must worry that when $\theta_0 \approx \pi$ the dynamics may be
rather more complicated than we have suggested. That is, rather than the
hot `normal' component providing a barrier between the interior and
exterior until the time of hadronization, if the energy in the interface
is large enough, the interface might plausibly play a role toward
`polarizing' the hadronization of the putative normal component.

This issue may become a little more clearly drawn in a slightly different
point of view (due to Bill Walker) regarding the mechanism for the
formation of the disoriented chiral condensates\refmark\refbjfour.
The picture is easiest
to describe in the rest frame of one of the colliding particles.  In this
frame, the oncoming particle is essentially a black disk as far as colored
objects are concerned.  (This picture is essentially due to the explosive
growth of small-x gluons at large momenta.)  As such, it acts as a
`vacuum cleaner', sweeping away all colored degrees of freedom associated
with the target.  What is left behind is `nothing', but not vacuum.  Since
it will be a color singlet, it will not become attached to the disk, but
will drift behind it, separated by a rapidity gap from the rest of the
collision debris.  This picture is essentially an alternative
description of the same initial stage of the collision which we described
in the opening section.  As such, we would expect the `nothing' to be
characterized by a total energy and momentum, and to expand outward in
the fashion described above, with a thin shell of `nothing' (hot partonic
matter) separating the rapidly cooling interior from the exterior.  But in
this picture, it is clear that as the `nothing' cools below the relevant
transition temperature, it is appropriate to describe it in terms of the
sigma model.  Thus, the distinction between the condensate and the `normal'
component is perhaps rather more blurred than we have been assuming,
particularly for events in which $\theta_0 \approx \pi$.

Finally, we turn to the question of estimating cross sections for events
such as we have been describing.
This is exceedingly difficult in the absence of a more detailed
understanding of the mechanism for the formation of the objects.  Bjorken
has suggested an approach to this question \refmark\refbjfour.
Roughly speaking, we know
that in order to have a `thermalized' state of rest energy $M^*$, we need
to have at least something of the order of
$M^*/m_0$ strongly interacting partons suitably localized.
($m_0$ is a QCD scale parameter
$\sim 1\; GeV$.)  Then the question becomes whether or not this is realized
in hadron-hadron collisions of a given energy.  This depends critically
on the small-x behaviour of the structure functions; the
QCD-inspired model  of Block, Halzen and
Margolis\Ref\bhm{M. M. Block, F. Halzen and B. Margolis,
{\sl Phys. Rev.} {\bf D45} 839 (1992).}
 simulates this behavior, and indicates that the bound is
probably easily satisfied at SSC energies.
However, it is not yet clear that the energy is deposited in a suitable
configuration.  A simple geometric argument based on the constituent quark
model suggests a lower bound of approximately a nanobarn, but this is
probably rather conservative, at least at supercollider energies.

\chapter{Are Centauros Related to Disoriented Chiral Condensates?}

Bjorken has suggested that the Centauro phenomena might be interpreted in
terms of disoriented chiral condensates\refbjone.
We now turn to this question.

Centauro events are cosmic ray events exhibiting\Ref\centauro{
C. M. G. Lattes, Y. Fujimoto and S. Hasegawa, {\sl Phys. Rep.} {\bf 65}
151 (1980).}:

\item{a.} Large ($ \sim 100$) numbers of hadrons;
\item{b.} Little apparent electromagnetic energy  and hence, no
$\pi^0$'s;
\item{c.} High hadronic $p_t$, reported as $k_\gamma \langle p_t \rangle
= 0.35 \pm .15\;GeV$, where $k_\gamma$ is the photon elasticity;

In addition to the Centauro events, a class of hadron-enriched events has also
been reported\Ref\cp{L.T. Baradzei, et al., {\sl Nucl. Phys.} {\bf B370}
365 (1992).}.
Cosmic ray events with $\sum E_{tot} \geq 100\; TeV$
are presented on a scatter plot of the number of hadrons $N_h$ versus the
fraction $Q_h = \sum E_h^{\gamma} /(\sum E_h^{\gamma} + \sum E_\gamma)$
of the visible energy which these hadrons constitute.  When compared
with Monte-Carlo simulations of families based on models of the strong
interactions, and assuming that cosmic ray primaries are predominantly
protons, there are far too many ($\sim 20\%$) events in regions not populated
by the Monte-Carlo.  These events show fluctuations in hadron number and/or
energy fraction.

Before continuing it is probably necessary to briefly review the
status of candidate Centauro events.  The 5 `classic' Centauro events
were seen in the two-storeyed emulsion chamber experiment of the
Brasil-Japan collaboration, located at 5220 m at the Chacaltaya
observatory in Bolivia.  At least two additional candidate Centauro events
have also been seen in Chacaltaya chambers.  At least one additional
candidate has been seen in the Pamir emulsion chamber experiment.  On
the other hand, it is claimed that Centauros have not been observed
in emulsion chambers at Mt. Kanbala (5500 m, China-Japan Collaboration) or
at Mt. Fuji (3750 m., Mt. Fuji Collaboration), despite comparable
cumulative exposures\Ref\fuji{J. R. Ren, et al., {\sl Phys. Rev.}
{\bf D 38} 1417 (1988).}.  More precisely, the China-Japan collaboration
reports an upper limit of the fraction of such events among hadron
families with energy greater than 100 TeV to be 3\% at the 95\%
confidence level.  This appears to be a limit incompatible with the
rate at which Chacatalya has observed Centauros. This comparison may
be too glib, however, because of differences in emulsion chamber
design and data analysis. Of particular importance may be the
differing techniques for
separating hadronic showers from others.

The two groups are
similarly divided on the (non)observation of the more general class of
hadron-enriched events mentioned above.  In this context,
the debate (which has been going on for over a decade) seems to reduce
to a question of whether the data signal a change in the composition of
cosmic ray primaries at these energies, or whether they signal a change
in the hadronic interactions.

With these caveats, we proceed to discuss the consistency of the Centauro
 and hadron-enriched phenomena with the phenomenology of chiral condensates
as developed in the preceeding sections.

We begin with the characteristic feature of Centauro's and the hadron-
enriched events:  anomalously large amounts of energy in the hadron
component.  The suppression of $\pi^0$'s which this implies is often
taken to indicate a suppression of pions altogether.  The argument is
basically statistical:  one would ordinarily expect the neutral fraction
to be given by essentially a binomial distribution, resulting in events
sharply peaked about $1/3$.  As we have seen above, however, the distribution
for an isospin-zero coherent state of pions is $\sim 1/\sqrt{f}$, with
$f\sim 0$ being the most probable fraction. (Note, however,
that $\langle f \rangle
=1/3$.)  Thus, one is tempted to interpret the classic Centauro events as
signals of a disoriented chiral condensate.

A problem immediately arises.  While the multiplicity of the Centauro
events is comparable to our estimates, we also suggested that a large
fraction of the hadrons would be `normal.'  This would appear to rule out
Centauro events as signals of a disoriented condensate.  The only apparent
resolution to this difficulty in interpretation is that these events represent
a situation for which our understanding of the interface between the
interior and the exterior at time of hadronization is suspect:  $\theta_0
\approx \pi$.

Assuming this to be true, we are faced with an additional problem: the
claimed anomalously high $\langle p_t\rangle$ of the Centauros.
This value depends on
the fact that Centauro I was close enough to the detector that the
position of the iteraction vertex could be determined by the angular
divergence of the showers in the detector.  The actual value measured is
$${\langle E_h^{(\gamma)} R_h\rangle\over H} = k_\gamma \langle p_t \rangle
=0.35 \pm 0.15 \; GeV,$$
where $E_h^{\gamma}$ is the portion of an incident hadron energy which
is converted to (visible) electromagnetic energy, $R_h$ is the distance
of an incident hadron from the center of the event, and $H$ is the
height of production.  In order to determine $p_t$, however, one needs
to know the value of the gamma-ray inelasticity, $k_\gamma$.
The range is usually quoted as
$0.2-0.4$, with the lower range being preferred for nucleons, while the
higher end is preferred for pions.  Direct measurement of $k_\gamma$ in
emulsion chambers is impossible because of the high energy threshold
($\sim 1\; TeV$).  As a result, estimates are based on extrapolating
accelerator data or on Monte-Carlo simulations.  These seem to indicate
that one should use $k_\gamma \sim 0.4$ or larger\Ref\kgamma{Chacaltaya
Collaboration,
``Gamma-ray Inelasticity $k_\gamma$ in Emulsion Chamber Experiment'',
in proceedings of
{\sl 22nd International Cosmic Ray Conference} (Dublin, Ireland) v. 4,
 p. 89 (1991).}
.  We would thus
estimate that $\langle p_t \rangle \sim 0.875 \pm 0.375\; GeV$
with large systematic
uncertainty. This seems compatible with our estimates of the previous
section.  We note most analyses have followed the Japan-Brazil collaboration
and have used $k_\gamma = 0.2$, based on the assumption that the hadrons
are nucleons.

We seem to be on somewhat firmer ground in confronting the hadron-enriched
events.  Detailed comparison will require rather careful Monte-Carlo
simulations, but the general features of the plot of $N_h$ versus
$Q_h$ seems to be entirely consistent with a phenomenology based on
disoriented condensates, with corresponding fluctuations in the neutral
fraction, and in $\theta_0$.

We are thus rather encouraged that the Centauro and the
hadron-enriched events may be signals of the creation of disoriented
chiral condensates.

Finally, if we accept this interpretation, then we can use the
experimental data to suggest both the cross section and the threshold
energy for the production of disordered chiral condensates.  As far
as the cross section is concerned, the abundance of the hadron-enriched
events suggest that they are almost generic at these energies (cosmic
ray showers with $>100\; TeV$ of visible energy).  We were unable to
track down more detailed estimates of the energies in the typical
hadron-enriched events, but one can determine a lower bound on the
threshold energy for Centauros.  In the 5 classic events, the estimated
visible energy of the showers (after correcting for atmospheric
effects) was $\sim 350 \; TeV$.  Dividing by $k_\gamma \sim 0.4$ yields an
estimated total energy of $875 \; TeV$.  If the parent interaction
is due to a collision between nucleons, then we can estimate the minimum
center of mass energy of the collision as $\sqrt{s}\sim 1.3 \;TeV$.  This
suggests that there is no inconsistency with the negative results
of the UA1 and UA5 Centauro searches\REFS\uaone{G. Arnison, et al.,
{\sl Phys. Lett.} {\bf 122B} 189 (1983).}
\REFSCON\uafive{G. J. Alner et al., {\sl Phys. Lett.} {\bf 180B} (1986).}
\REFSCON\uafivepr{G. J. Alner, et al., {\sl Phys. Rep.} {\bf 154} 247 (1987).}
\refsend.

\chapter{Experimental Challenges}

We conclude that there is a considerable body of evidence
based on high-energy cosmic-ray
observations,
buttressed by some theoretical considerations,
that there may be a significant fraction of
the pp and p$\bar p$ inelastic cross section
above
$\sqrt{s} \sim
1.3\; TeV$ that have 'anomalous' pionic multiplicity patterns.
We have described
one model for accounting for such multiplicity patterns above a certain
threshold energy.  The Tevatron energy is above this threshold and lies in
the lower range of the spectrum of anomalous cosmic-ray events; the LHC and
the SSC are well above the anticipated threshold energy and cover the
spectrum of most of the cosmic-ray events.

	The characteristics of these events and their interpretion have been a
subject of widespread interest in the cosmic-ray, particle-physics, and
astrophysics communities for over a decade.  This continued interest
represents considerable justification for continuing their experimental
investigation at the Tevatron, the LHC, and the SSC. The replication of the
below-threshold UA1 and UA5 minimum-bias experiments at the Tevatron would
constitute a natural bridge for similar experiments at super-collider
energies and would replace the  lower-energy cosmic-ray data, whose sample
size is very limited and whose processing is both complicated and model
dependent, by much more extensive experimental data not subject to the same
procedural uncertainties.  These data would provide extremely valuable
guidance for the choice and refinement of the theoretical modeling of the
production and symmetry-breaking mechanisms of strong-interaction physics
that involve complicated collective effects rather than better understood
perturbative processes.  This guidance is likely to be more uniequivocal,
at least in regard to chiral-symmetry breaking mechanisms, than corresponding
investigations at RHIC where large nuclear/deconfinement effects are likely
to lead to much more complicated signals.

	In summary, collider experiments at the Tevatron, the LHC, and the SSC
looking for generic, high-threshold, unusual-multipicity-pattern
hadron-hadron inelastic collisions are needed for the following reasons:

\item{(1)} The intrinsic interest of the continued investigation of unusual
strong interaction production mechanisms at Tevatron (and higher) energies.

\item{(2)}
 To replace most of the limited-sample-size cosmic-ray data set (which is
based on a variety of observational setups, models, and assumptions) by the
results of controlled experiments.

\item{(3)} To provide the experimental input that is needed to discriminate
between the large number of theoretical conjectures that have been proposed
to account for the cosmic-ray results as well as those that relate to the
symmetry-breaking mechanisms of QCD.  Experimental guidance is urgently
needed for the modeling of the complicated collective effects that seem to
be responsible for physics of this kind.

\refout
\end

\end